\documentclass[prl,twocolumn,aps]{revtex4}

\usepackage{graphicx} 

\newcommand{\ket}[1]{\left|#1\right\rangle}

\begin{document}

\title{Analytic asymptotic performance of topological codes}

\author{Austin G. Fowler}
\affiliation{Centre for Quantum Computation and Communication
Technology, School of Physics, The University of Melbourne, Victoria
3010, Australia}

\date{\today}

\begin{abstract}
Topological quantum error correction codes are extremely practical, typically requiring only a 2-D lattice of qubits with tunable nearest neighbor interactions yet tolerating high physical error rates $p$. It is computationally expensive to simulate the performance of such codes at low $p$, yet this is a regime we wish to study as low physical error rates lead to low qubit overhead. We present a very general method of analytically estimating the low $p$ performance of the most promising class of topological codes. Our method can handle arbitrary periodic quantum circuits implementing the error detection associated with this class of codes, and arbitrary Pauli error models for each type of quantum gate. Our analytic expressions take only seconds to obtain, versus hundreds of hours to perform equivalent low $p$ simulations.
\end{abstract}

\maketitle

A number of proposals exist for a 2-D array of qubits with tunable nearest neighbor interactions \cite{Devi08,Amin10,Jone10,Kump11}. Any 2-D topological quantum error correction code \cite{Brav98,Bomb06,Ohze09b,Katz10,Bomb10} and some 3-D codes \cite{Raus07,Fowl09} can be mapped with little or no overhead to such hardware. Despite the undemanding physical requirements, most forms of topological quantum error correction (TQEC) possess very high threshold error rates. The threshold error rate of a quantum error correction code is the physical error rate $p_{\rm th}$ below which it is possible to perform arbitrarily reliable quantum computation. The best TQEC schemes possess threshold error rates of order 1\% \cite{Wang11,Fowl11b}.

\begin{figure}
\begin{center}
\resizebox{55mm}{!}{\includegraphics{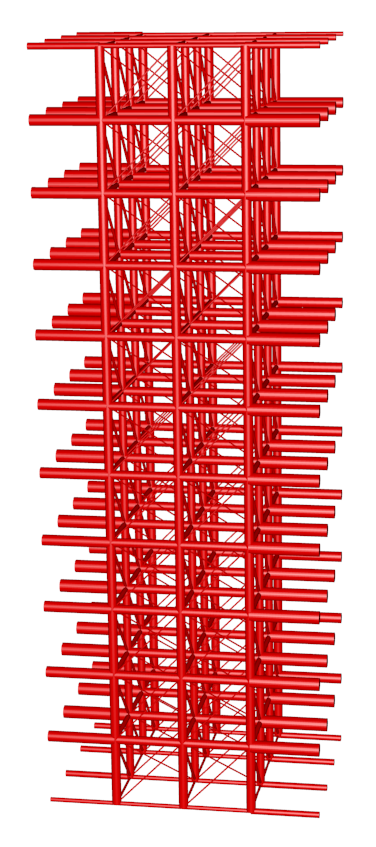}}
\end{center}
\caption{An example of a surface code nest. Each cylinder represents the total probability of a single error leading to detection events at the endpoints of the cylinder. Time runs vertically.}\label{nest}
\end{figure}

Viewing the 2-D array of qubits together with the duration of a quantum computation as a 3-D volume, we define a TQEC circuit to be any 3-D periodic quantum circuit that detects errors (using combinations of single-qubit measurement results) in such a way that detection events can be unambiguously divided into two classes, primal and dual, and a single physical error away from boundaries of the 3-D volume always leads to 0 or 2 local primal and dual detection events. Since distinct detection events are associated with distinct sets of qubit measurements, and measurements begin at specific space-time locations, every detection event can be systematically associated with a unique point in space-time. We can use our tool Autotune \cite{Fowl12d} to efficiently determine the total probability of single errors leading to a pair of (primal or dual) detection events at any chosen pair of space-time locations. We visualize each such probability as a cylinder between the relevant space-time locations with diameter proportional to the probability. We call each cylinder a stick, and call any large collection of sticks a nest. Fig.~\ref{nest} contains an example of a (primal) nest generated by Autotune. This nest is associated with the surface code \cite{Denn02,Raus07d,Fowl08}. Only a basic understanding of the definitions of detection events and nests contained in this paragraph is required to read this work.

The distance $d$ of a TQEC instance can be determined from its nests by counting the minimum number of sticks required to connect distinct boundaries. Boundaries in fig.~\ref{nest} are indicated by sticks apparently leading to nowhere, so it can be seen that this example has distance $d=4$. By analyzing a particular nest in the limit of large $d$, we have been able to prove \cite{Fowl12e} that the surface code possesses a non-zero threshold error rate when minimum weight perfect matching \cite{Edmo65a,Edmo65b,Fowl12c} is used to correct errors. In this work we extend of the techniques of \cite{Fowl12e} much further, and describe how to automatically calculate two analytic expressions $p^{\rm pr}_L(d,p)$ and $p^{\rm du}_L(d,p)$, valid for even $d$, for the asymptotic low $p$ performance of any TQEC circuit as defined above with any Pauli error model for each type of gate. Our new method requires seconds of computation versus typically hundreds of hours to do an equivalent simulation based low $p$ analysis of a TQEC circuit.

We begin with a few definitions. A logical operator is any path of sticks connecting distinct boundaries. A fault is any selection of $d/2$ sticks from a logical operator. The probability of a logical operator $p_{\rm op}$ or fault $p_{\rm f}$ is the product of probabilities of its constituent sticks. We define the compliment probability $p_{\rm c}$ to be $p_{\rm op}/p_{\rm f}$, namely the product of the probabilities of the sticks in the logical operator that are not in the fault. Note that more than one logical operator can contain the same fault. Minimum weight perfect matching based TQEC fails when a fault occurs that is contained in a logical operator such that $p_{\rm c}>p_{\rm f}$. The matching algorithm will insert corrections corresponding to the most probable selection of sticks that generates the observed detection events, which will result in a logical operator that corrupts the stored quantum data. Failure will also occur 50\% of the time when $p_{\rm c}=p_{\rm f}$ as this case corresponds to two different equally probable complementary faults that lead to the same detection events. Given the identical detection events for both faults, matching will randomly succeed 50\% of the time.

Each layer of boundary sticks in fig.~\ref{nest} corresponds to the execution of a single round of error detection circuitry --- a complete period of the circuit. Our goal is to formally define and determine the total probability of all faults causing failure per round of error detection.

The first step towards our goal is to imagine we have a nest of infinite temporal extent and to choose a particular round of error detection to focus on. More specifically, we wish to focus on the boundary sticks associated with a single boundary in that round. In practice, we generate nests with $2d+4$ rounds, as shown in fig.~\ref{nest}, and choose round $d+2$ --- this ensures that everything we consider is well away from temporal boundaries.

Armed with our chosen boundary sticks, we sequentially generate every $d$-stick logical operator starting on each of these boundaries sticks and sequentially generate every fault contained in each logical operator. Every time we generate a new fault, we add it to a set, including its value of $p_{\rm c}$ relative to the current logical operator. If the fault is already present, $p_{\rm c}$ is adjusted to the maximum of the old and new values. This is done to make sure that only the most likely fault compliments are considered. When complete, we will have a base set of faults.

We need a set of faults uniquely associated with each round of error detection. Our base set is too large, containing many faults that should be associated with earlier or later rounds. We therefore consider the boundary sticks one round higher, round $d+3$, and again sequentially generate every $d$-stick logical operator and fault. If a fault is not found in our base set, we add it to a step set, adjusting $p_{\rm c}$ if required as described above if the fault is already in step. If a fault is found in our base set and the new $p_{\rm c}$ is higher, the fault is moved to the step set with the new $p_{\rm c}$. The step set constructed in this manner does not contain faults that should be associated with rounds in the past. It will, however, contain faults that should be associated with rounds in the future.

We continue considering successively higher rounds of boundary sticks, sequentially generating all logical operators and faults until we fail to generate a single fault matching a fault in step but with higher $p_{\rm c}$. When this process is complete, our step set will contain only faults best associated (highest $p_{\rm c}$) with logical operators making use of one of the $d+3$ round boundary sticks. We can now sum the fault probabilities $p_{\rm f}$, using a scale factor of 0 if $p_{\rm f}>p_{\rm c}$, 0.5 if $p_{\rm f}=p_{\rm c}$, or 1 if $p_{\rm f}<p_{\rm c}$.

\begin{figure}
\begin{center}
\resizebox{85mm}{!}{\includegraphics[viewport=0 60 650 430, clip=true]{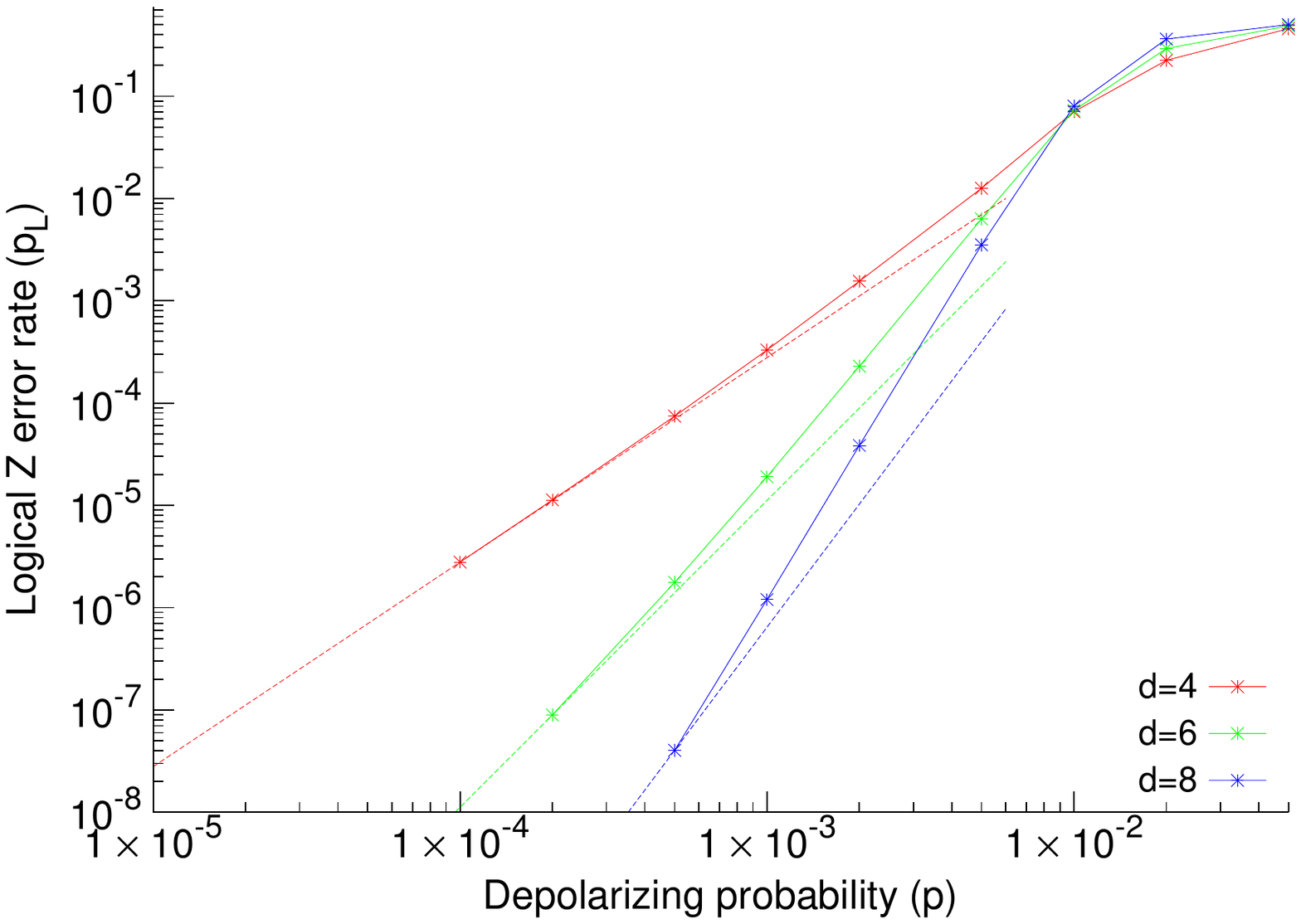}}
\resizebox{85mm}{!}{\includegraphics[viewport=0 60 650 430, clip=true]{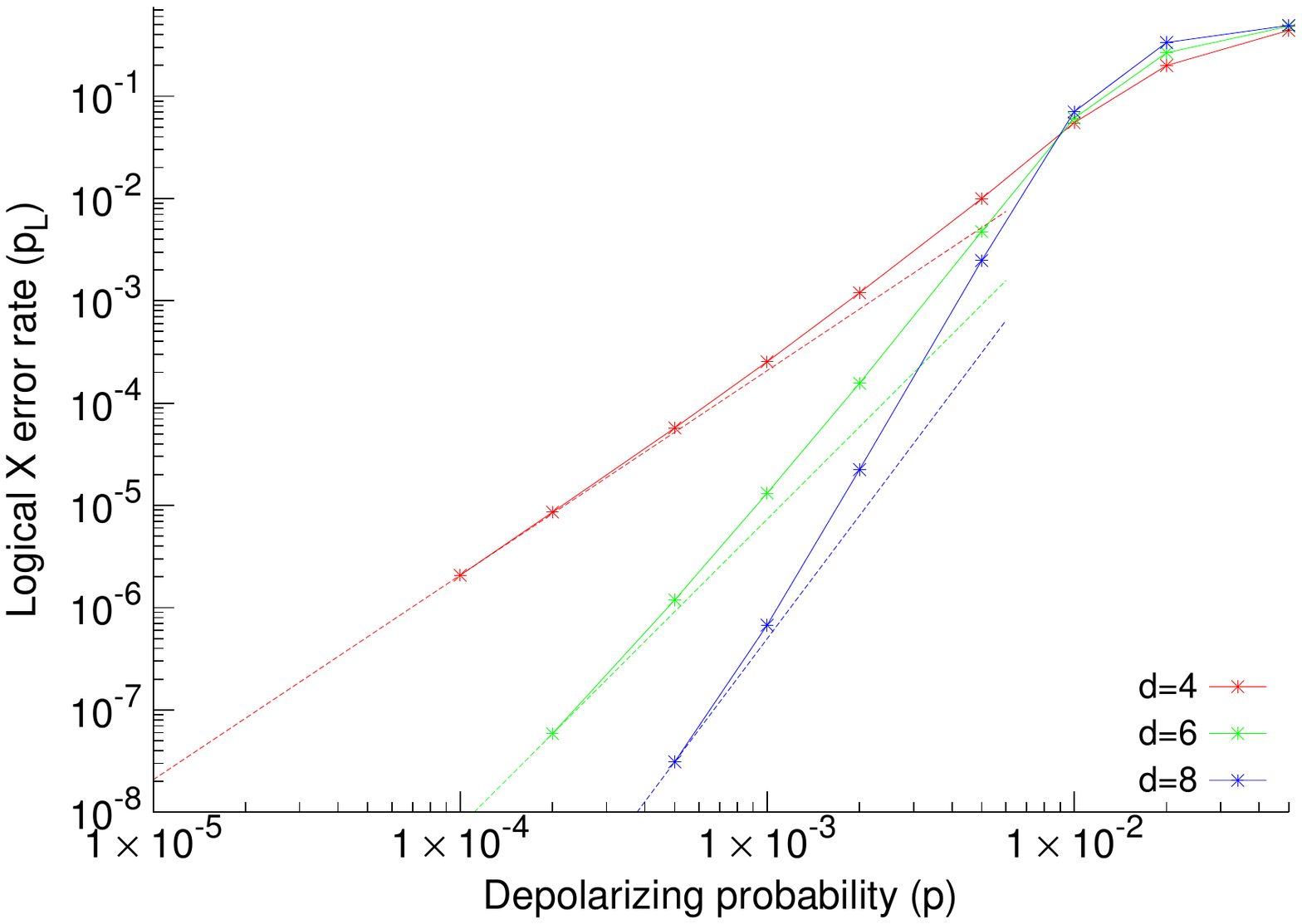}}
\end{center}
\caption{Probability of (top) logical $Z$ (primal) and (bottom) logical $X$ (dual) error per round of error correction for various surface code distances $d$ and physical error rates $p$. The asymptotic curves (dashed lines) are quadratic, cubic and quartic.}\label{logicalzx}
\end{figure}

Having described our method, we turn our attention to its performance. Fig.~\ref{logicalzx} show simulation data for a standard surface code making use of initialization to $\ket{0}$ and $\ket{+}$ and measurement in the $Z$ and $X$ bases each with probability $p$ of failure, CNOT and identity gates with depolarizing noise of probability $p$, and all gates of equal duration. These seemingly innocuous curves took over 1500 CPU hours to generate, with the vast majority of the time spent generating the lowest $p$ data points. Despite the substantial computational effort, it can be seen that the $d=8$ curves in particular have not yet well converged to their expected low $p$ quartic asymptotic form (quartic due to any fault leading to failure involving a minimum of 4 sticks). The coefficients $A$ of each simulation determined $Ap^{d/2}$ can be found in Table~\ref{comp}, along with the time required to obtain the lowest $p$ data point. Note that our simulations include the full depolarizing channel and hence all of our simulations return both the primal and dual performance.

\begin{table}
\begin{tabular}{c|c|c|c|c|c|c}
$d$ & $A_{\rm pr}$ & $A_{\rm du}$ & $t_A$ & $B_{\rm pr}$ & $B_{\rm du}$ & $t_B$ \\
\hline
4 & $2.79\times 10^2$ & $2.08\times 10^2$ & 270hr & $2.90\times 10^2$ & $2.33\times 10^2$ & 0.5sec \\
6 & $1.12\times 10^4$ & $7.32\times 10^3$ & 480hr & $1.08\times 10^4$ & $7.38\times 10^3$ & 2sec \\
8 & $6.44\times 10^5$ & $4.98\times 10^5$ & 480hr & $3.99\times 10^5$ & $2.39\times 10^5$ & 7sec \\
10 & --- & --- & --- & $1.47\times 10^7$ & $7.73\times 10^6$ & 70sec \\
\end{tabular}
\caption{Surface code distance $d$, stochastic simulation determined primal and dual asymptotic coefficients ($Ap^{d/2}$), time required to obtain the lowest $p$ data points used to calculate the asymptotic coefficients, analytic primal and dual asymptotic coefficients ($Bp^{d/2}$), time required to perform the analytic computation.}
\label{comp}
\end{table}

Table~\ref{comp} also contains analytic coefficients $B$ obtained using our new method. For distances 4 and 6, where decent asymptotics could be obtained using stochastic simulations, the agreement is good. Discrepancies may be due to the assumption in the analytic calculation of independent stick probabilities, which are not independent in simulations. Given a Pauli error model is a significant approximation of the physics of real devices, the maximum 12\% discrepancy for distance 4 dual asymptotics is of no practical concern. Of great practical benefit is the ability to analytically estimate the asymptotics of distances out of reach of simulation. It can be seen from fig.~\ref{logicalzx} that the distance 8 asymptotic curve is expected to be lower, an expectation that is confirmed by the analytics in table~\ref{comp}. We can now claim a solid understanding of the error processes leading to failure in TQEC schemes of the type considered in this work.

From \cite{Fowl12e}, we expect exponential suppression of logical error with increasing $d$ at fixed $p$ below threshold. This expectation is supported by the ratio of successive pairs of primal $B$ coefficients being approximately 37 and that of the dual $B$ coefficients being approximately 32. This allows us to write $p^{\rm pr}_L(d,p)=0.21(37p)^{d/2}$ and $p^{\rm du}_L(d,p)=0.23(32p)^{d/2}$, valid for $p\lesssim 10^{-4}$. Note that only the distance 4 and 6 analytic data are required to construct these expressions, requiring 2.5 seconds of computation versus 750 hours to obtain the same data by simulation, a factor of a million improvement.

In summary, we have described a highly computationally efficient method of obtaining accurate analytic expressions describing the low $p$ performance of topological quantum error correction, the form of quantum error correction that leads to the lowest space and time overhead for large quantum computations and the parameter region that was previously computationally almost inaccessible. Our method can cope with the details of real hardware, including geometric constraints, gate duration asymmetries, gate error rate asymmetries and arbitrary Pauli error models for each type of gate as our method builds on our tool Autotune. We expect our method to facilitate the real-time interactive investigation of the dependence of the error correction performance on the details of the hardware, something impossible via simulation, greatly assisting the search for less experimentally challenging pathways towards large-scale commercially viable quantum computation.

This research was conducted by the Australian Research Council Centre of Excellence for Quantum Computation and Communication Technology (project number CE110001027), with support from the US National Security Agency and the US Army Research Office under contract number W911NF-08-1-0527. Supported by the Intelligence Advanced Research Projects Activity (IARPA) via Department of Interior National Business Center contract number D11PC20166.  The U.S. Government is authorized to reproduce and distribute reprints for Governmental purposes notwithstanding any copyright annotation thereon.  Disclaimer: The views and conclusions contained herein are those of the authors and should not be interpreted as necessarily representing the official policies or endorsements, either expressed or implied, of IARPA, DoI/NBC, or the U.S. Government.

\bibliography{../References}

\end{document}